\documentclass[secnumarabic,aps,prl,reprint,groupedaddress]{revtex4-1}

\usepackage[english]{babel}
\usepackage{graphicx}
\usepackage{color}
\usepackage{amsmath}
\usepackage{tabularx}

\begin{document}
	
\newcommand{\unit}[1]{\:\mathrm{#1}}            
\newcommand{\To}{\mathrm{T_0}}
\newcommand{\Tp}{\mathrm{T_+}}
\newcommand{\Tm}{\mathrm{T_-}}
\newcommand{\EST}{E_{\mathrm{ST}}}
\newcommand{\Rp}{\mathrm{R_{+}}}
\newcommand{\Rm}{\mathrm{R_{-}}}
\newcommand{\Rpp}{\mathrm{R_{++}}}
\newcommand{\Rmm}{\mathrm{R_{--}}}
\newcommand{\ddensity}[2]{\rho_{#1\,#2,#1\,#2}} 
\newcommand{\ket}[1]{\left| #1 \right>} 
\newcommand{\bra}[1]{\left< #1 \right|} 

\title{Energy and momentum conservation in spin transfer}
\author{Alexander Mitrofanov}
\author{Sergei Urazhdin}
\affiliation{Department of Physics, Emory University, Atlanta, GA, USA.}

\begin{abstract}
We utilize simulations of spin-polarized electron scattering by a chain of localized quantum spins to show that energy and linear momentum conservation laws impose strong constraints on the properties of magnetic excitations induced by spin transfer. In turn, electron's orbital and spin dynamics depends on the dynamical characteristics of the local spins. Our results suggest the possibility to achieve precise control of spin transfer-driven magnetization dynamics by tailoring the spectral characteristics of the magnetic systems and the driving electrons.
\end{abstract}

\maketitle

The advent of spin transfer (ST) effect~\cite{Slonczewski1996,Berger1996,PhysRevB.57.R3213} has transformed our understanding of nanomagnetism, and spurred multiple novel applications~\cite{Kent2015, Divinskiy2016, Locatelli2013,Kim2012,7505988}. ST is caused by the interaction of spin currents carried by conduction electrons with the magnetization of magnetic materials, resulting in the absorption of electron's spin angular momentum component non-collinear with the magnetization~\cite{Slonczewski1996,Zhang2002,Ralph20081190}. The absorbed angular momentum drives magnetization dynamics, which can result in magnetization reversal~\cite{PhysRevLett.84.3149,Mangin2006}, precession~\cite{Kiselev2003,PhysRevLett.92.027201} and other dynamical effects~\cite{Demidov2010,Madami2011}.

Energy and linear momentum conservation laws play a central role in the dynamical processes in nature, but their relevance to ST has remained virtually unexplored. The threshold current for ST-driven magnetization dynamics was initially attributed to the requirement that spin accumulation must exceed the energy $E_m$ of the magnetic excitation quanta [magnons] generated by ST~\cite{PhysRevLett.80.4281}. However, the energy of magnons associated with quasi-uniform magnetization precession excited by ST is small, and the threshold was identified with the compensation of the dynamical damping by ST~\cite{PhysRevLett.84.3149,Ralph20081190}.

Recent studies showed that ST can excite dynamical modes throughout the magnon spectrum~\cite{Lee2004,Polianski2004}, which spans frequencies $f_m$ from GHz to THz ranges for common ferromagnets (Fs)~\cite{PyDispertion}. Excitation of high-frequency magnons may play a significant role in the interplay between thermal phenomena and ST~\cite{Bauer2012}. Nonlinear interactions among these high-frequency modes can also profoundly influence ST-induced dynamics~\cite{Demidov2011,Divinskiy2019}. ST can also drive magnetic dynamics in antiferromagnets (AFs)~\cite{Jungwirth2016,elezn2018,Moriyama2018}, where the lowest dynamical frequencies are typically in 100s of GHz or in the THz range~\cite{RevModPhys.30.1,Kampfrath2010}, which may enable ultrafast devices and THz oscillators based on AFs driven by ST~\cite{Khymyn2017}.

The energies $E_m=hf_m$ of THz magnons are in the meV range. If energy conservation plays a role in ST, a large electrical bias may be required to provide energy sufficient to generate such magnons. Likewise, linear momentum conservation may impose strict requirements on the momentum of the driving electrons in magnetic nanodevices envisioned to operate with short-wavelength magnons generated by ST~\cite{Grundler2016,Divinskiy2016}. However, both energy and momentum of magnons have been neglected in the analyses of ST, which with a few exceptions~\cite{PhysRevB.69.134430,PhysRevB.85.092403,PhysRevLett.119.257201,PhysRevB.99.094431,PhysRevB.99.024434} have approximated magnetization as a classical vector field.

Here, we use simulations of spin-polarized electron scattering by a quantum spin chain to show that energy and momentum conservation laws impose significant constraints on the magnetic dynamics, as well as the electron's orbital and spin dynamics resulting from ST. Our results suggest the possibility to control the characteristics of magnetic excitations generated by ST by optimizing these constraints, which may provide a new route for the development of efficient magnetic nanodevices.

To analyze ST, we consider scattering of an electron wavepacket by a ferromagnet modeled as a 1D spin-1/2 chain. In the tight-binding approximation, this system can be described by the Hamiltonian~\cite{PhysRevB.99.094431,supplemental}
\begin{equation}\label{eq:hamiltonian}
\begin{split}
\hat{H}=-\sum_{i}b|i\rangle\langle i+1|\\
-\sum_{j}J_{sd}|j\rangle\langle j|\otimes \hat{\mathbf{S}}_j\cdot\hat{\mathbf{s}}+J\hat{\mathbf{S}}_j\cdot\hat{\mathbf{S}}_{j+1}+\mu_B\hat{S}_j^zB,
\end{split}
\end{equation}
where indices $i,j$ enumerate the tight-binding sites, $\hat{\mathbf{s}}$, $\hat{\mathbf{S}}_j$ are the spin operators of the electron and the local spins, $b$ is the electron hopping parameter, $J$ describes the exchange stiffness of the local spins, $J_{sd}$ - their exchange with the electron, $B=-B_z$ is the magnetic field, and $\mu_B$ is the Bohr magneton. We use periodic boundary conditions for both the electron and the spin chain, to avoid spurious effects of reflections at the boundaries.

\begin{figure}
	\includegraphics[width=\columnwidth]{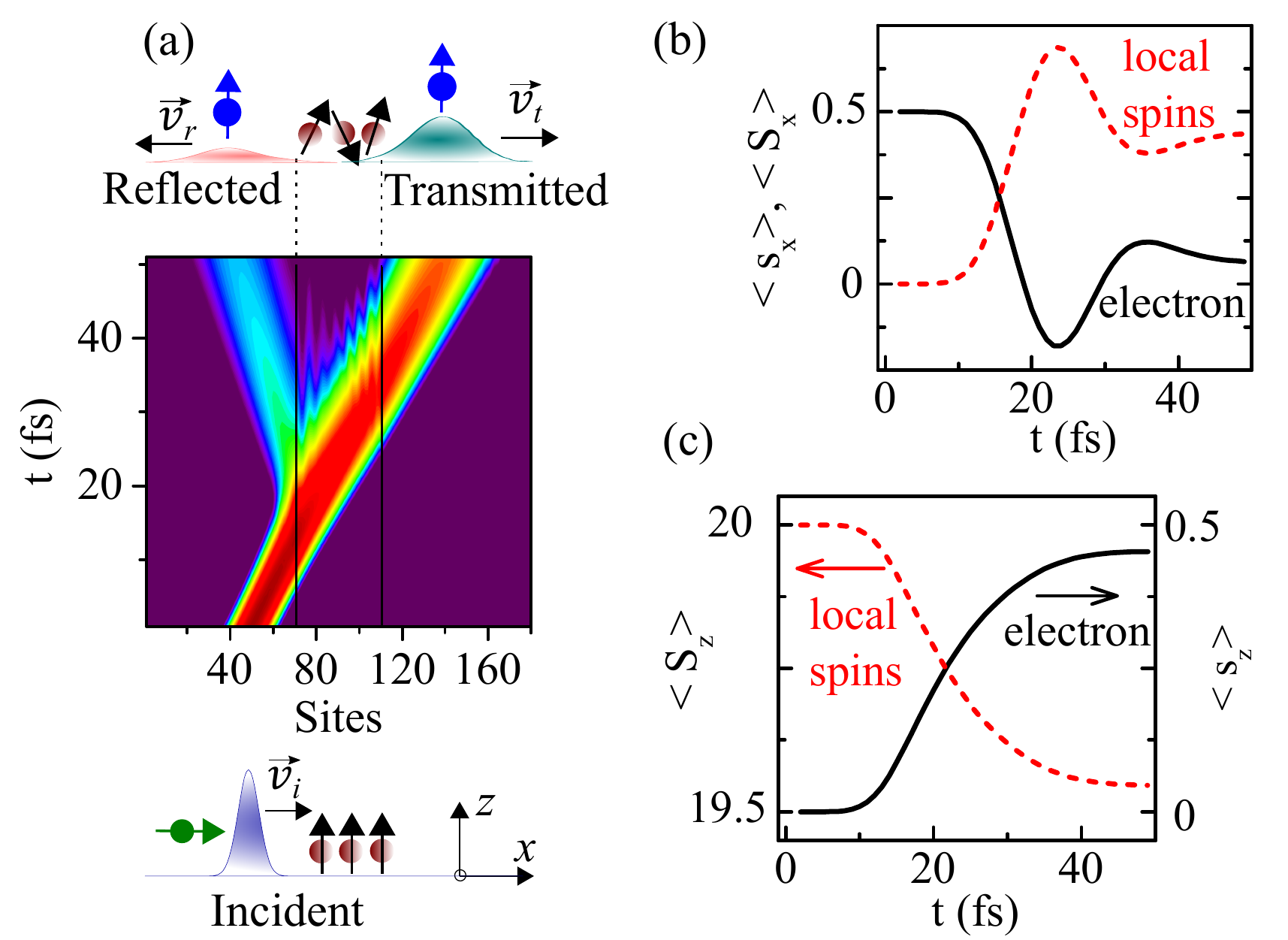}	
	\caption{\label{fig:qst_intro} (Color online) ST due to scattering of electron wavepacket by the chain of $40$ spins, with , $a=0.2$~nm, $b=1$~eV, $J=J_{sd}=0.1$~eV, $B=20$~T. (a) Pseudocolor map of wavepacket intensity in the position-time coordinates, for the initial wavepacket polarization along the x-axis. Schematics: the wavepacket and the spin chain before and after scattering. (b),(c) Evolution of the expectation values of x (b) and z (c) components of electron's and chain's spins.}
\end{figure}

To analyze ST, the system is initialized with the electron forming a Gaussian wave packet spin-polarized along the x-axis, while the local spins are in their ground state aligned with the z-axis. The system is then evolved according to the Hamiltonian Eq.~(\ref{eq:hamiltonian}). The wavepacket is partially reflected and partially transmitted by the local spins [Fig.~\ref{fig:qst_intro}(a)]. One can clearly identify the time intervals when the wavepacket is localized mostly outside or inside the spin chain, allowing us to analyze the effects of scattering by tracking the time evolution. 

To analyze the evolution of each subsystem, we introduce the density matrices $\hat{\rho}_e=Tr_m\hat{\rho}$ and $\hat{\rho}_m=Tr_e\hat{\rho}$ for the electron and the local spins, respectively, by tracing out the full density matrix $\hat{\rho}$ with respect to the other subsystem~\cite{PhysRevB.99.094431}. The expectation value of an observable $\hat{A}$ associated with the electron is $\left<\hat{A}\right>=Tr(\hat{A}\hat{\rho}_e)$, while the probability of its value $a$ is $P_a=\left<\psi_a|\hat{\rho}_e|\psi_a\right>$, where $\psi_a$ is the corresponding eigenstate. Similar relations hold for the observables associated with the local spins.

Exchange interaction of the electron with the local spins results in the oscillation of its x spin component, which rapidly decays due to dephasing [Fig.~\ref{fig:qst_intro}(b)], consistent with the ST mechanisms~\cite{Slonczewski1996,Ralph20081190}. The x-component of the local spins mirrors this evolution, so that the x-component of the total spin is conserved. The contribution of the Zeeman term in Eq.~(\ref{eq:hamiltonian}) that breaks the spin conservation is negligible on the considered time scales.

The z-component of electron spin increases from zero to almost its maximum value $1/2$, with the local spins mirroring this evolution, Fig.~\ref{fig:qst_intro}(c). This transfer of the spin component collinear with the magnetization is consistent with the recently demonstrated nonclassical contribution to ST~\cite{PhysRevLett.119.257201,PhysRevB.99.094431,supplemental}. Since the constraints imposed by energy and momentum conservation are expected to be general, we do not separate between the two contributions to ST in the analysis below.

\begin{figure}
	\includegraphics[width=\columnwidth]{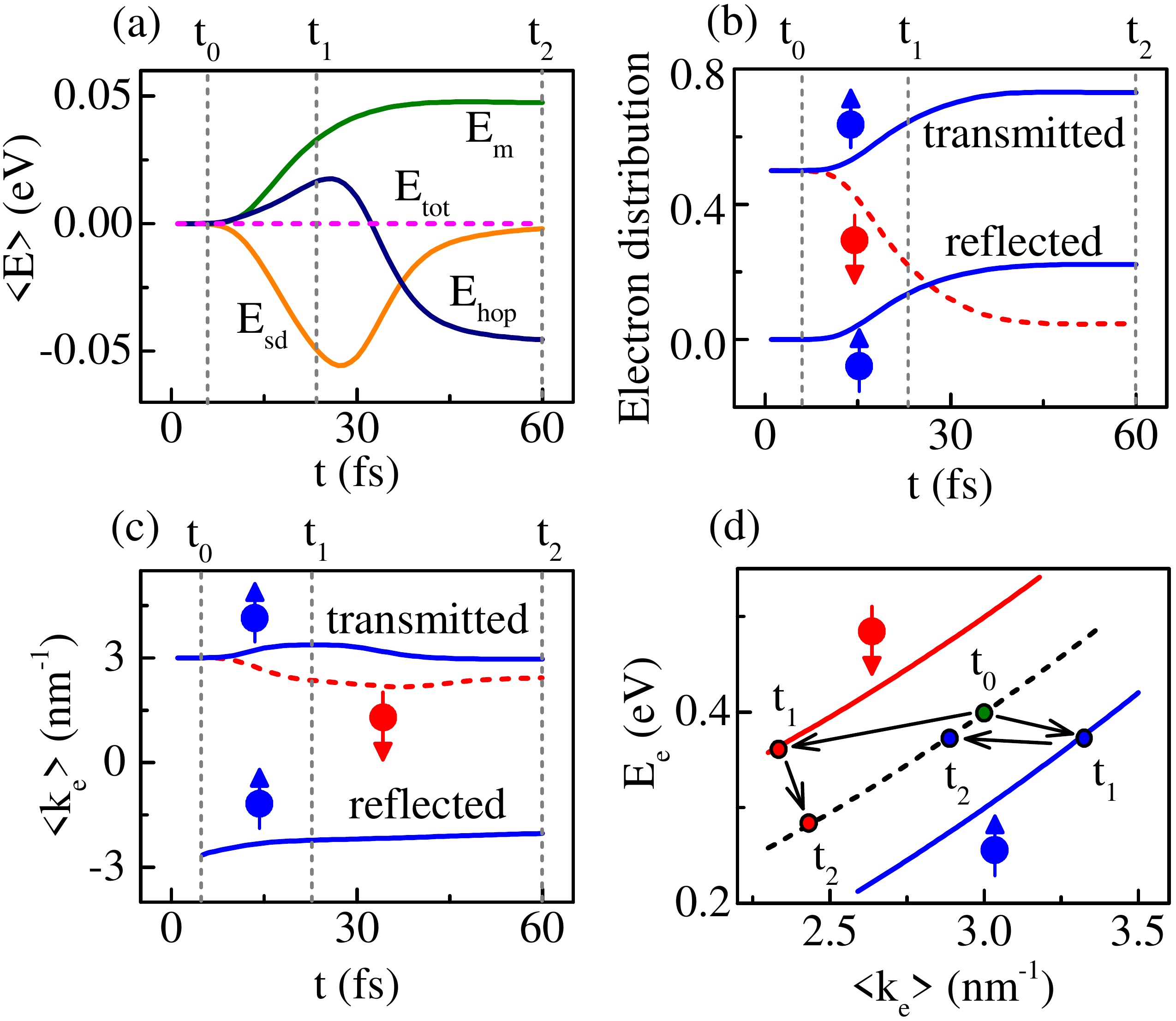}	
	\caption{\label{fig:en_mom} (Color online) (a) Evolution of different contributions to energy, as defined in the text. The curves are shifted by the $t=0$ values for clarity. (b),(c) Majority (solid curves) and minority (dashed) contributions to the transmitted and reflected wavepacket components (b), and the corresponding average wavevectors (c). The reflected minority component (not shown) is negligible. (d) Energy vs momentum for forward-propagating electron components at times $t_0$, $t_1$, and $t_2$, as marked in (a),(b). Solid curves: spin-dependent electron dispersion inside the spin chain, dashed curve - dispersion outside the chain.}
\end{figure}

The evolution of different contributions to energy is illustrated in Fig.~\ref{fig:en_mom}(a). The magnetic energy $E_m$ comprising the Zeeman and the exchange energies of the local spins increases due to their excitation by ST, while the exchange energy $E_{sd}$ between the local spins and the electron initially decreases due to the increase of the electron's spin-up [majority] component [see Fig.~\ref{fig:qst_intro}(c)]. The two subsystems no longer interact after scattering, so $E_{sd}$ increases back to zero. Since the Hamiltonian is time-independent, the total energy of the system is conserved [dashed line in Fig.~\ref{fig:en_mom}(a)]. The deficit of energy associated with a finite $E_m$ after scattering is made up by the reduction of the electron's kinetic energy $E_{hop}$. This suggests that the relation between the electron's kinetic energy and the magnetic excitation spectrum plays an important role in ST, as confirmed below.

We now analyze the momentum evolution. Before scattering, the wave packet contains only the forward-propagating component, with equal majority and minority spin contributions, Fig.~\ref{fig:en_mom}(b). During scattering, the minority contribution decreases, while the majority contribution increases, consistent with the transfer of z spin component shown in Fig.~\ref{fig:qst_intro}(c). Additionally, a majority-spin backward-propagating component emerges due to the electron reflection by the spin chain. The reflected minority-spin component is negligible in the approximation of the same electron hopping parameter inside and outside the spin chain, consistent with the mechanisms of electron-magnon scattering discussed below. 

The momentum of the reflected majority-spin component is considerably smaller than that of the original wave packet [Fig.~\ref{fig:en_mom}(c)], indicating that electron reflection by the chain involves a large transfer of energy. Meanwhile, the momentum of the majority-spin forward-propagating component increases, and that of the minority component decreases as the electron enters the chain, consistent with the spin-splitting of the electronic band structure inside the chain due to the sd exchange [see Fig.~\ref{fig:en_mom}(d)]. However, the difference between the momenta of the two spin components remains significant even after scattering, indicating that the electron experiences spin-dependent momentum and energy loss. This is confirmed by Fig.~\ref{fig:en_mom}(d), which shows the average momentum and energy of the forward-propagating components calculated for instants $t_0$, $t_1$, and $t_2$ before, during, and after scattering, as marked in panels (a) and (b). The momentum and the energy of the majority-spin component are slightly reduced at $t_2$ relative to $t_0$, while those of the minority-spin component are significantly reduced.

\begin{figure}
	\includegraphics[width=\columnwidth]{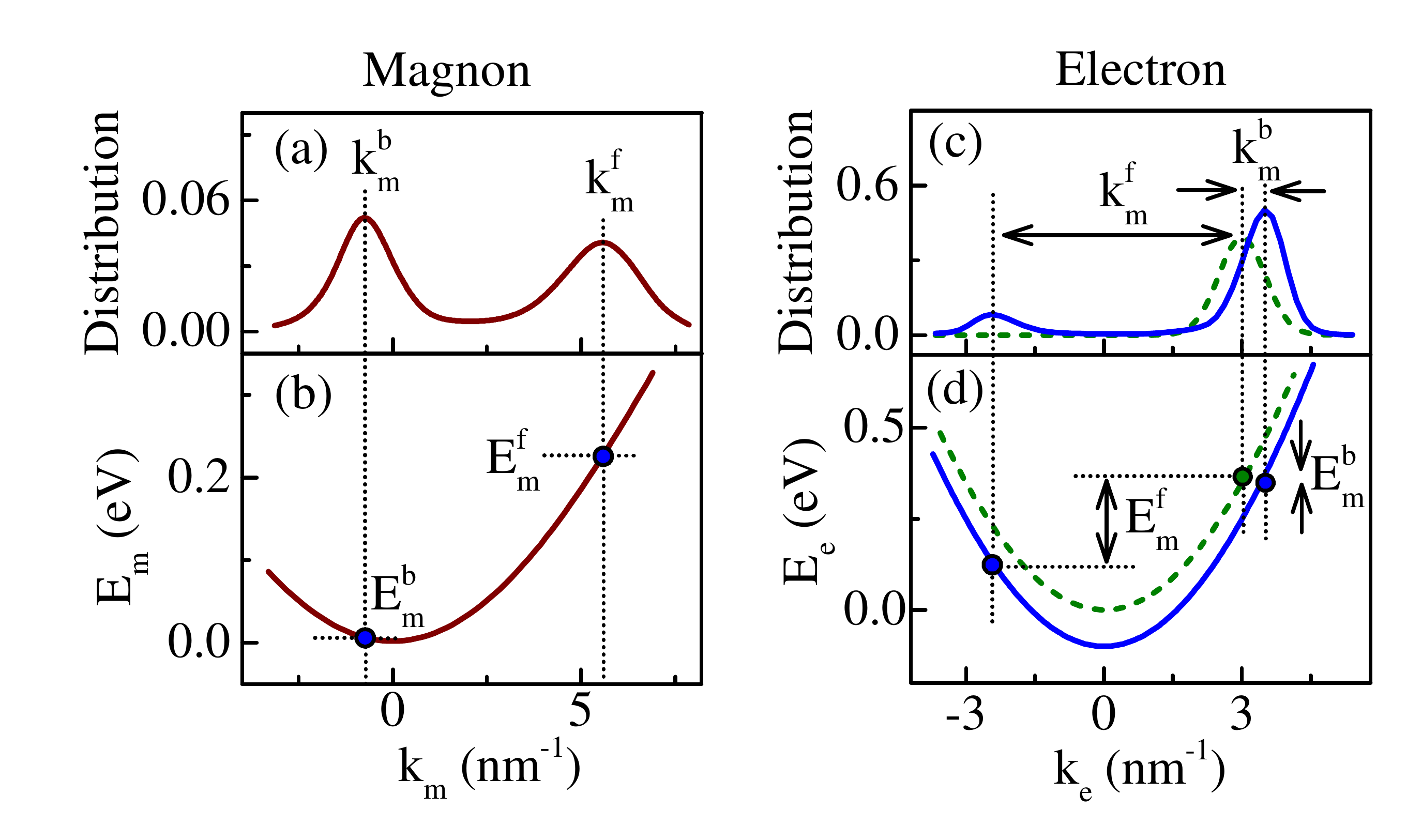}	
	\caption{\label{fig:el_mag} (Color online) (a) Momentum distribution of the generated magnons, at $t=t_1$. (b) Magnon dispersion. (c) Momentum distribution of the majority-spin wave packet component at $t=t_0$ (dashed curve) and at $t=t_1$ (solid curve). (d) Electron dispersion $E_e=2b(1-\cos(k_ea))$ outside the spin chain (dashed curve), and majority-spin  dispersion $E_{e,\uparrow}=2b(1-\cos(k_ea))-J_{sd}$ in the spin chain (solid curve). The momenta $k^f_m$, $k^b_m$ and the energies $E^f_m$, $E^b_m$ of the forward- and the backward-propagating magnon groups are indicated. 
	} 
\end{figure}

The variations of the electron's energy and momentum [Fig.~\ref{fig:en_mom}] are inconsistent with quasi-elastic scattering, suggesting that the energy and the momentum of the generated magnons play a significant role in the scattering process. This is confirmed by the analysis of the relation between the  distribution of the generated magnons and the characteristics of the wave packet, Fig.~\ref{fig:el_mag}. Two distinct groups of magnons are generated: forward-propagating magnons with a large central  momentum $k^f_m$, and backward-propagating magnons with a small centeral momentum  $k^b_m$ [Fig.~\ref{fig:el_mag}(a)]. We use the magnon dispersion relations $E_m=4J(1-\cos(k_ma)))+S\mu_B B$, where $a$ is the tight-binding site spacing, to determine the corresponding magnon energies $E^f_m$ and $E^b_m$ [Fig.~\ref{fig:el_mag}(b)].

The relations between the momenta of the generated magnons and the characteristics of the electron wavepacket are illustrated in Fig.~\ref{fig:el_mag}(c), which shows the majority-spin momentum distributions of the wave packet at $t=t_0$ and at $t_1$. The difference between the initial central momentum $k^i_e$ of the wave packet and the momentum $k^r_e$ of the reflected component is equal to the momentum $k^f_m$ of the forward-propagating magnons generated due to ST, while the corresponding difference for the momentum $k^t_e$ of the transmitted wave packet component is equal to the momentum $k^b_m$ of the generated backward-propagating magnons. 

By analyzing the dispersion of the electron outside the spin chain, as well as the majority-spin dispersion of electron inside the spin chain [Fig.~\ref{fig:el_mag}(d)], we find that the energy $E^r_e$ of the reflected component is reduced relative to the initial energy $E^i_e$ by the energy $E^f_m$ of the forward-propagating magnons generated by scattering, while the energy $E^t_e$ of the transmitted component is reduced by the energy $E^b_m$ of the backward-propagating magnons. Here, the term ``energy" refers to the expectation value of energy of the corresponding quantum-mechanical state, rather than the net energy carried by the wave. Thus, generation of forward-propagating magnons is associated with electron reflection, while generation of backward-propagating magnons - with the forward scattering of electrons, described by the relations
		\begin{equation}
		\label{eq:en_mom}
		E_{e}^i=E_{m}^{f(b)}+E_{e}^{r(t)},\;\;
		k_{e}^i=k_{m}^{f(b)}+k_{e}^{r(t)}
		\end{equation}
between the energies and the momenta of the quasiparticles involved in the corresponding scattering processes. To confirm our interpretation, we solved these equations using the magnon and the electron dispersions. For instance, the equation for the momentum $k^t_e$ of the transmitted electron is
\begin{equation}\label{eq:decision}
b-\frac{S\mu_B B-J_{sd} + 4J\cos(k_{e}^ia-k_{e}^ta)}{2(\cos(k_{e}^ta)-\cos(k_{e}^ia))}=0.
\end{equation}
Its numeric solution is consistent with Fig.~\ref{fig:el_mag}(c)~\cite{supplemental}.

Equation~(\ref{eq:en_mom}a) describes energy conservation, as expected for the time-independent Hamiltonian Eq.~(\ref{eq:hamiltonian}). However, its translation symmetry is broken by the spin chain, so the momentum needs not be conserved. Indeed, the momentum of the forward-propagating majority electron becomes reduced after scattering, as expected since its energy is reduced due to magnon generation [see Fig.~\ref{fig:en_mom}(d)]. However, the generated magnon with momentum $k^b_m$ propagates backward, i.e. the total momentum is reduced in this process. Nevertheless, the momentum relation Eq.~(\ref{eq:en_mom}b) is governed by the same spatial interference between the spin wave and the incident/scattered electron wavefunctions as in the momentum-conserving processes, and therefore we for simplicity call it the momentum conservation condition.

\begin{figure}
	\includegraphics[width=\columnwidth]{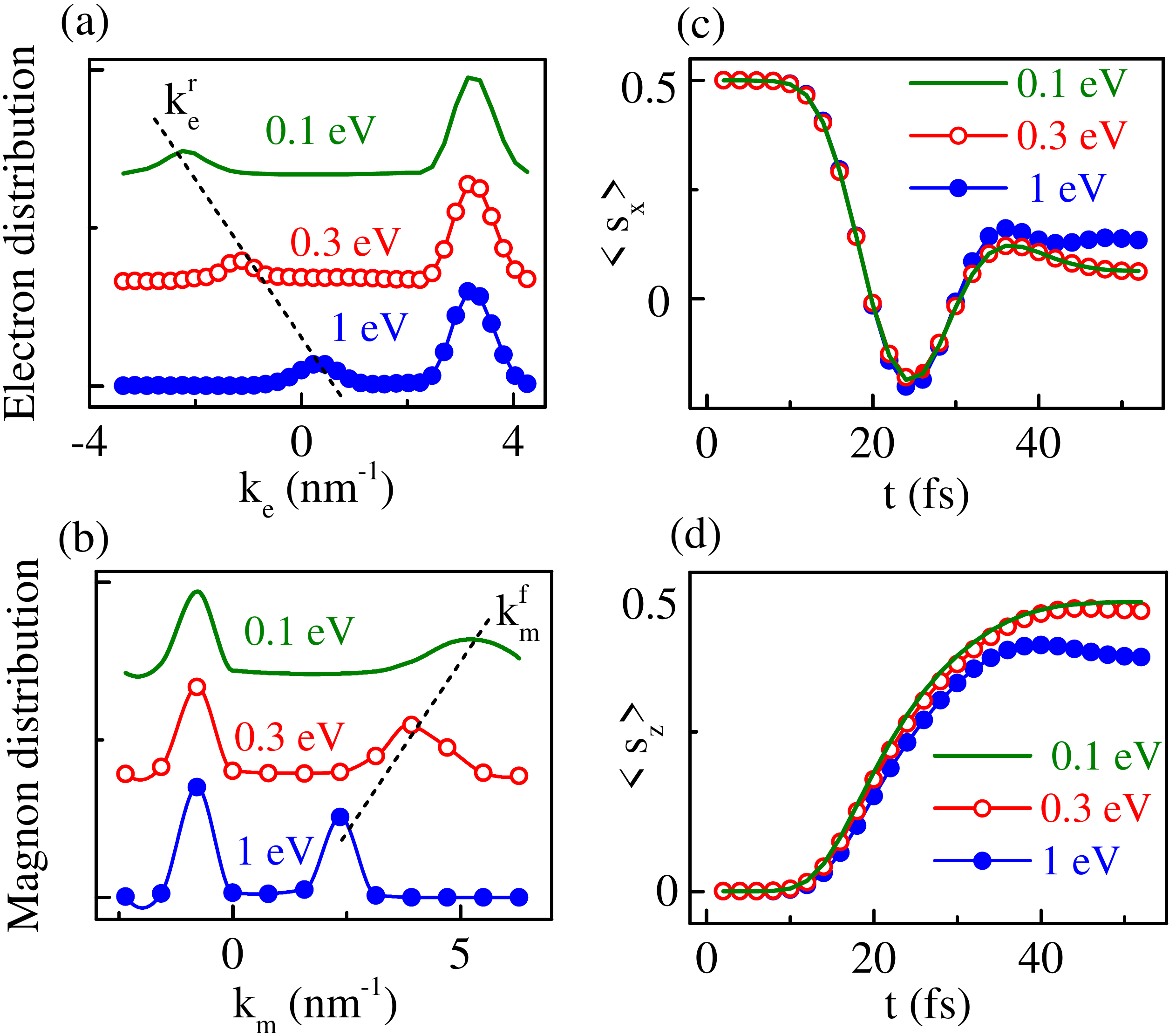}	
	\caption{\label{fig:ex_stif} (Color online) Effects of magnon dispersion on ST. (a),(b) Electron (a) and magnon (b) momentum distributions at $t=t_1$, for the labeled  values of $J$. (c),(d) Evolution of the x- (c) and the z-component of electron spin (d).}
\end{figure}

Electron scattering described by Eq.~(\ref{eq:en_mom}) is governed by the electron and magnon dispersions. Here, we demonstrate one of the consequences - dependence of electron scattering and ST on the magnon dispersion - which is not captured by the models based on the classical approximation for magnetization~\cite{supplemental}. 

Figure~\ref{fig:ex_stif}(a) shows the electron momentum distributions at the instant $t_1$, for three different values of exchange stiffness $J$. The transmitted component is not significantly affected by the variations of $J$, as expected since the energy of the magnons with a small momentum $k^b_m$, which are generated by the transmitted electrons, is almost independent of the exchange stiffness. In contrast, the magnitude of the momentum of the reflected component rapidly decreases with increasing $J$, which is mirrored by the decrease of the momentum $k^f_m$ of the generated magnons [Fig.~\ref{fig:ex_stif}(b)]. This effect is consistent with the increase of the energy $E^f_m$ of these large-momentum magnons, resulting in a decrease of the scattered electron's energy. At $J=1$~eV, the momentum of the scattered electron becomes close to zero, i.e. all of its initial energy is transferred to the generated magnon.

The evolution of both the x- and the z-components of the electron spin is similar for $J=0.1$~eV and $J=0.3$~eV, [Figs.~\ref{fig:ex_stif}(c),(d)]. However, for $J=1$~eV, the transfer of both the x- and the z- components of spin is reduced. The electron's energy is no longer sufficient to generate the largest-momentum magnons, resulting in a reduced efficiency of ST. In our simulation, electron can be scattered into any band states, so this effect of magnon dispersion on ST becomes noticeable only at large $J$, when the magnon energies become comparable to the electron band energy. In real systems, the available electron energy is much smaller, as defined by the occupied Fermi surface. Consequently, a significant dependence of electron scattering and spin dynamics on the magnon dispersion can be expected even for modest variations of $J$ or other parameters controlling the magnon dispersion, such as the magnetic anisotropy or field~\cite{supplemental}. We leave analysis of these effects to future studies.

To summarize, we have shown that energy and momentum conservation laws define the energies and the momenta of magnons generated in the spin transfer process. As one of the consequences, the spectral distribution of spin waves generated by spin transfer in tunnel junctions must significantly differ from those in metallic systems. The demonstrated relations may provide a path for the development of laser-like magnetic nanodevices, where specific magnetic modes are excited by spin transfer due to the judicious optimization of constraints imposed by the conservation laws.

The demonstrated relations are relevant not only to spin transfer, but also to orbital and the spin dynamics of electrons scattered by the ferromagnets. For instance, electron backscattering at magnetic interfaces, which involves generation of large-momentum magnons, should strongly depend on the available electron energy. The constraints imposed on spin transfer by the conservation laws are also particularly relevant for antiferromagnets, where the characteristic magnon energies are two orders of magnitude larger than in ferromagnets. 

This work was supported by the U.S. Department of Energy, Office of Science,
Basic Energy Sciences, under Award \# DE-SC2218976.

\bibliography{QST_sim}
\bibliographystyle{apsrev4-1}

\end{document}